\documentclass[twocolumn]{aastex62}

\newcommand{\HI}{H\,\textsc{i}}
\newcommand{\SigTP}{\Sigma_{\rm TP}}

\graphicspath{{./}{figures/}}

\shorttitle{Auddy et al.}
\shortauthors{Auddy, Basu, and Kudoh}
\begin{document}

%\title{The Transition from a Lognormal to a Power-law Column Density Distribution in Molecular Clouds: An Imprint of the Initial Magnetic Field and Turbulence}
\title{THE TRANSITION FROM A LOGNORMAL TO A POWER-LAW COLUMN DENSITY DISTRIBUTION IN MOLECULAR CLOUDS: AN IMPRINT OF THE INITIAL MAGNETIC FIELD AND TURBULENCE}

\correspondingauthor{Sayantan Auddy}
\email{sauddy@asiaa.sinica.edu.tw}

\author[0000-0003-3784-8913]{Sayantan Auddy}
%\author{Sayantan Auddy}
\affil{Institute of Astronomy and Astrophysics, Academia Sinica, Taipei 10617, Taiwan}

\author[0000-0003-0855-350X]{Shantanu Basu}
\affiliation{Department of Physics and Astronomy, The University of Western Ontario, London, ON N6A 3K7, Canada}
% \email{basu@uwo.ca}

\author{Takahiro Kudoh}
\affiliation{Faculty of Education, Nagasaki University, 1-14 Bunkyo-machi, Nagasaki 852-8521, Japan}
% \email{kudoh@nagasaki-u.ac.jp}
% \collaboration{(AAS Journals Data Scientists collaboration)}

% \author{Butler Burton}
% \affiliation{National Radio Astronomy Observatory}
% \affiliation{AAS Journals Associate Editor-in-Chief}
% \nocollaboration

% \author{Amy Hendrickson}
% \altaffiliation{Creator of AASTeX v6.2}
% \affiliation{TeXnology Inc.}
% \collaboration{(LaTeX collaboration)}

% \author{Julie Steffen}
% \affiliation{AAS Director of Publishing}
% \affiliation{American Astronomical Society \\
% 2000 Florida Ave., NW, Suite 300 \\
% Washington, DC 20009-1231, USA}

% \author{Jeff Lewandowski}
% \affiliation{IOP Senior Publisher for the AAS Journals}
% \affiliation{IOP Publishing, Washington, DC 20005}



\begin{abstract}
% We show that in star-forming molecular clouds, both initial magnetic field and turbulence influence the value of the transitional column density between the lognormal and the power-law form of the probability distribution function (PDF).

% Turbulent magnetohydrodynamic simulations of star-forming molecular clouds shows that both initial magnetic field and turbulence influence the value of the transitional column density between the lognormal and the power-law form of the probability distribution function (PDF). We develop an analytic expression for $\sigma_{\rm TP}$ based on the interplay of turbulence, strong magnetic field and gravity. Our expression for $\sigma_{\rm TP}$

	We introduce a theory for the development of a transitional column density $\Sigma_{\rm TP}$ between the lognormal and the power-law forms of the probability distribution function (PDF) in a molecular cloud. Our turbulent magnetohydrodynamic simulations show that the value of $\Sigma_{\rm TP}$ increases as the strength of both the initial magnetic field and turbulence increases. We develop an analytic expression for $\Sigma_{\rm TP}$ based on the interplay of turbulence, a (strong) magnetic field, and gravity. The transition value $\Sigma_{\rm TP}$ scales with $\mathcal{M}^2_{\rm 0}$, the square of the initial sonic Mach number, and $\beta_{0}$, the initial ratio of gas pressure to magnetic pressure. We fit the variation of $\Sigma_{\rm TP}$ among different model clouds as a function of $\mathcal{M}^2_{\rm 0} \beta_{0}$, or equivalently the square of the initial Alfv\'enic Mach number $\mathcal{M}^2_{\rm A0}$. This implies that the transition value $\Sigma_{\rm TP}$ is an imprint of cloud initial conditions and is set by turbulent compression of a magnetic cloud. Physically, the value of $\Sigma_{\rm TP}$ denotes the boundary above which the mass-to-flux ratio becomes supercritical and gravity drives the evolution. 
%We point out how observations of the column density PDF and a measured value of the sonic Mach number can be used to estimate the large-scale mass-to-flux ratio of a molecular cloud.

\end{abstract}

\keywords{ISM: clouds--- magnetic fields--- magnetohydrodynamics (MHD) --- 
stars: formation }


\section{Introduction}\label{sec:intro}
% Understanding Star formation:  The theory of star formation has evolved from the standard paradigm of ``strong magnetic field" governed star formation to ''turbulence (week magnetic field)" model or '' turbulence-enhanced-ambipolar-diffusion" models. However, ``How stars are formed?" and in particular "how magnetic field affects star formation" is still open problem primarily due to lack of observational constraints.
% The theory of star formation holds a position of fundamental importance towards understanding formation of large scale structure like galaxy to much smaller scale like planet formation. However, In spite of decades of research.  The formation of star is a complex process governed by the interplay of gravity, turbulence and magnetic field ().   ... (TBA and edited)   

The column density probability distribution function (PDF) provides an effective way to analyze the dynamics and the evolution of molecular clouds from both observational \citep[e.g.,][]{bur15b,sch15b,sch16,pok16} and theoretical \citep[e.g.,][]{bur18,bas19} perspectives. Numerical simulations have established that non-self-gravitating gas with driven turbulence results in many interacting shocks that yield a lognormal density or column density PDF \citep[e.g.,][]{pad97,pas98,sca98,fed08,mol12}. The addition of self-gravity into simulations introduces a high-density power-law tail to the PDF \citep{kri11,par11,col12,fed13,war14,aud18}. Observed column density PDFs have an underlying lognormal shape with an additional power-law tail \citep{kai09,alv14} that starts at a transitional column density \citep{sch15}.
% The column density probability distribution function (PDF) of molecular clouds has played a pivotal role in understanding star formation. Thus this naturally raises the question: can these differences in the column density structure of molecular clouds be used to distinguish the relative importance of turbulence, gravity and magnetic field?
% The question is are there differences in the column density structure of molecular clouds that can distinguish the relative importance of turbulence, gravity and magnetic field? to explain initial mass function \citep{hen08}, star formation rate \citep{fed12} and star formation efficiency \citep{fed13} of molecular clouds. 

A lognormal shape is associated with quiescent clouds that do not have active star 
formation \citep{kai09,lom15,sch15}. In contrast, the active star-forming clouds have an excess of high column density with a prominent power-law tail as well as a lognormal peak. The lognormal feature is often considered to be a direct imprint of driven supersonic turbulence and its width is attributed to the strength of the sonic Mach number \citep{col12,mol12,bur15}. However, it has also been suggested that the lognormal may be a more general characteristic that is set by both supersonic turbulence and gravitationally-driven ambipolar diffusion \citep{tas10} or global gravitational contraction \citep{par11}. 

%Furthermore, recent observations by \cite{alv17} suggest that lognormal peak may be an artefact arising due to data incompleteness, and thereby not a result of supersonic turbulence.

% Width maps to Mach number: The popular interpretation is that the lognormal feature is a direct imprint supersonic turbulence and its width is related the sonic Mach number \citep{col12}(burkhart et al 2015). It could be a more generic characteristic that influenced by gravitational driven ambipolar diffusion \cite{tas10} and/or global gravitational contraction \citep{par11} along with supersonic turbulence. However, recent observations by \cite{alv17} claims that lognormal peak may be an artefact arising due to data incompleteness, and thereby not a result of supersonic turbulence.

% The power-law tail is an effect of growth in high column density regions due to gravitational contraction. But variation in the steepness of the slope can be a consequence of initial field strength. Increased magnetic field strength can cause weak steepening of the power-law tail \citep{col12}
The power-law part of the PDF \citep[][]{sch13,alv17} is a signature of gravitational contraction (due to the self-gravity of the gas) and can be associated with the formation of condensed cores. The PDF $dN/d\log \Sigma \propto \Sigma^{-\alpha}$ has an index $\alpha =2$ 
in the limit of isothermal gravitational contraction \citep[see e.g., Appendix A of][]{aud18}, and is set by the density profiles within dense cores. However, the observed $\alpha$ is sometimes steeper. For instance, molecular clouds like Polaris and Pipe have power-law indices $\alpha = 3.9$ and $\alpha = 3.0$, respectively \citep{lom15}. These clouds are diffuse and have much less star formation compared to active star-forming clouds like Aquila which have $ \alpha \approx 2$ \citep{kon15}.
\citet{aud18} showed that the magnetic field can significantly affect the slope of the power-law tail.
% Numerical simulations of turbulence accelerated strongly magnetic medium \citep[][]{aud18} show that magnetic field can influence the slope of the power-law tail and affect the star formation rate. 
Clouds with a strong magnetic field (subcritical mass-to-flux ratio) and small amplitude initial perturbations develop a steep power-law tail ($\alpha \approx 4$), consistent with gravitationally-driven ambipolar diffusion leading to shallower core density profiles
than in a hydrodynamic collapse. In contrast, turbulent subcritical clouds retain the lognormal shape for a long time and eventually develop a power-law tail with $\alpha \approx 2$ in a region that has become supercritical due to turbulence-enhanced ambipolar
diffusion. 

Each PDF has at least three measurable parameters: the width of the lognormal part, the slope of the power-law tail, and the transitional column density $\SigTP$ that separates the lognormal from the power-law portion. Many theoretical studies have associated the standard
deviation $\sigma$ of the lognormal distribution with the sonic Mach number of driven
turbulence \citep[e.g.,][]{fed08,mol12}. The power-law tail develops when self-gravity is introduced into a driven turbulence simulation \citep{col12,fed13}. 
%, sometimes accomplished by fiat by turning on self-gravity
%when the cloud has reached a steady-state of turbulence through driving with Fourier space perturbations. 
Decaying hydrodynamic turbulence simulations with self-gravity also show a rapid development of a power-law tail in the PDF \citep{kri11,par11,war14}. 
Previous studies have not developed a theory for the location of $\SigTP$, although
\citet{bur17} and \citet{ima16} have proposed that it is associated with the 
\HI-to-H$_2$ transition in the interstellar medium. This has some appeal
since \HI\, clouds are known to be non-self-gravitating whereas molecular clouds 
exist at higher pressures and are considered to be self-gravitating \citep{bli91}.
However, observations of many molecular clouds show that the transition occurs 
within the molecular gas, and that
the value of $\Sigma_{\rm {TP}}$ is unique to each cloud, e.g., NGC 3603, Carina, Maddalena, and Auriga all have different deviation points (DP) as listed in Table 1 of \citet{sch15}. This implies that $\Sigma_{\rm {TP}}$ is an imprint of initial conditions inherent to a particular star-forming cloud and is set by physical processes.
Furthermore, the longstanding well-known low efficiency of star formation \citep[e.g.,][]{gol08} within molecular clouds means that most of the molecular gas mass is 
not within gravitationally-contracting dense cores that account for the
power-law portion of the PDF.

%These attributes have been linked with important aspects of star formation theory, e.g., the kinematics of an isothermal cloud \citep{fed08}, the initial mass function \citep{pad02,hen08}, the star formation rate \citep{kru05,pad11,hen11}, the star formation efficiency \citep{fed13} and more recently the \HI-to-H$_2$ transition \citep{bur17,ima16,bia17}. 
%The simulation papers in the above list work in the paradigm of driven
% turbulence in molecular clouds that is implemented through continuous perturbations in Fourier space. Self-gravity is initially turned off, and then switched
% on after a turbulent steady-state has been developed. A power-law tail rapidly develops in the pdf once self-gravity is turned on, since the clouds have supercritcal mass-to-flux ratio.
 
In this Letter, we focus on the physical origin of the transition point
$\SigTP$, using a different approach than adopted in most previous 
studies of the PDF. \citet{aud18} showed that in a decaying turbulence 
scenario with supercritical mass-to-flux ratio, the lognormal body is quickly
lost and a power law is developed for essentially all densities past the peak. 
Simulations with constant turbulent driving with Fourier space perturbations are able to
maintain a distinct lognormal body and a power law tail in the PDF, although
that may be an artefact of turning on self-gravity only after a steady turbulent
driving has been established. 
Since real molecular clouds do not have a ``switch-on'' gravity, we investigate
here the scenario that a strong magnetic field (i.e., subcritical
mass-to-flux ratio) supports large amplitude oscillations (even while there
is an overall decay of turbulence) that result
in the maintenance of a lognormal-like body of the PDF. Gravity is always at
work, but can only win out in dense regions that have undergone a rapid 
turbulence-accelerated ambipolar diffusion. In this case,
\citet{aud18} showed that a power-law tail with $\alpha \approx 2$ is added
to the lognormal-like body of the PDF. Here, we develop a self-consistent 
theory of the origin of $\SigTP$ in this scenario, and test it against a
suite of simulations with different initial conditions. We find a direct
link between $\SigTP$ and the relative importance of turbulence and 
magnetic fields in the initial cloud.
In Section \ref{sec:Numsim}, we present numerical simulations that study the properties of the column density PDFs. In Section \ref{sec:analytic} we derive an analytic expression for $\SigTP$ based on a model of turbulent compression of a magnetized cloud, and compare with the numerical results. In Sections \ref{sec:discussion} and \ref{sec:conclusion} we discuss and summarize our results, respectively.

\section{Column Density PDFs}\label{sec:Numsim}
% \subsection{Numerical Parameters and Setup} 
%  We solve the three-dimensional non-ideal MHD equations including self-gravity and ambipolar diffusion. We use a similar numerical setup as was previously used in \cite{kud07,kud11} and \cite{aud18}. The only modification in our current simulations is that we have doubled the resolution in both the $x-$ and $-y$ direction. The number of grid points in each direction is $(N_{x},N_{y}, N_{z}) = (512,512,20)$. 

%  The initial magnetic field is uniform and oriented along the $z-$ direction. The self-gravitating cloud is stratified along the direction of the magnetic field. It settles in to a hydrostatic equilibrium to form a sheet-like geometry. We use random velocity perturbations ($v_{x} = v_{a}R_{m}(x,y)$, $v_{y} = v_{a}R_{m}(x,y)$, $v_{z}$=0 ) at each grid points (where $R_{m}$ is random number with a spectrum $v_{k}^{2} \propto k^{-4}$ in the Fourier space) as a proxy for turbulence. The turbulence is allowed to decay freely and is not replenished. For more details about the set up of the simulation box, initial conditions and boundary conditions please refer to our previous paper \cite[][hereafter ABK18]{aud18}. The only modification in our current simulations is that we have doubled the resolution in both the $x-$ and $-y$ direction. The number of grid points in each direction is $(N_{x},N_{y}, N_{z}) = (512,512,20)$.

% \subsection{Simulated Column Density PFDs}

\begin{figure*}
\begin{center}
	\includegraphics[height=10cm,width=7.2in,trim=0mm 0mm 0mm 0mm, clip=true]{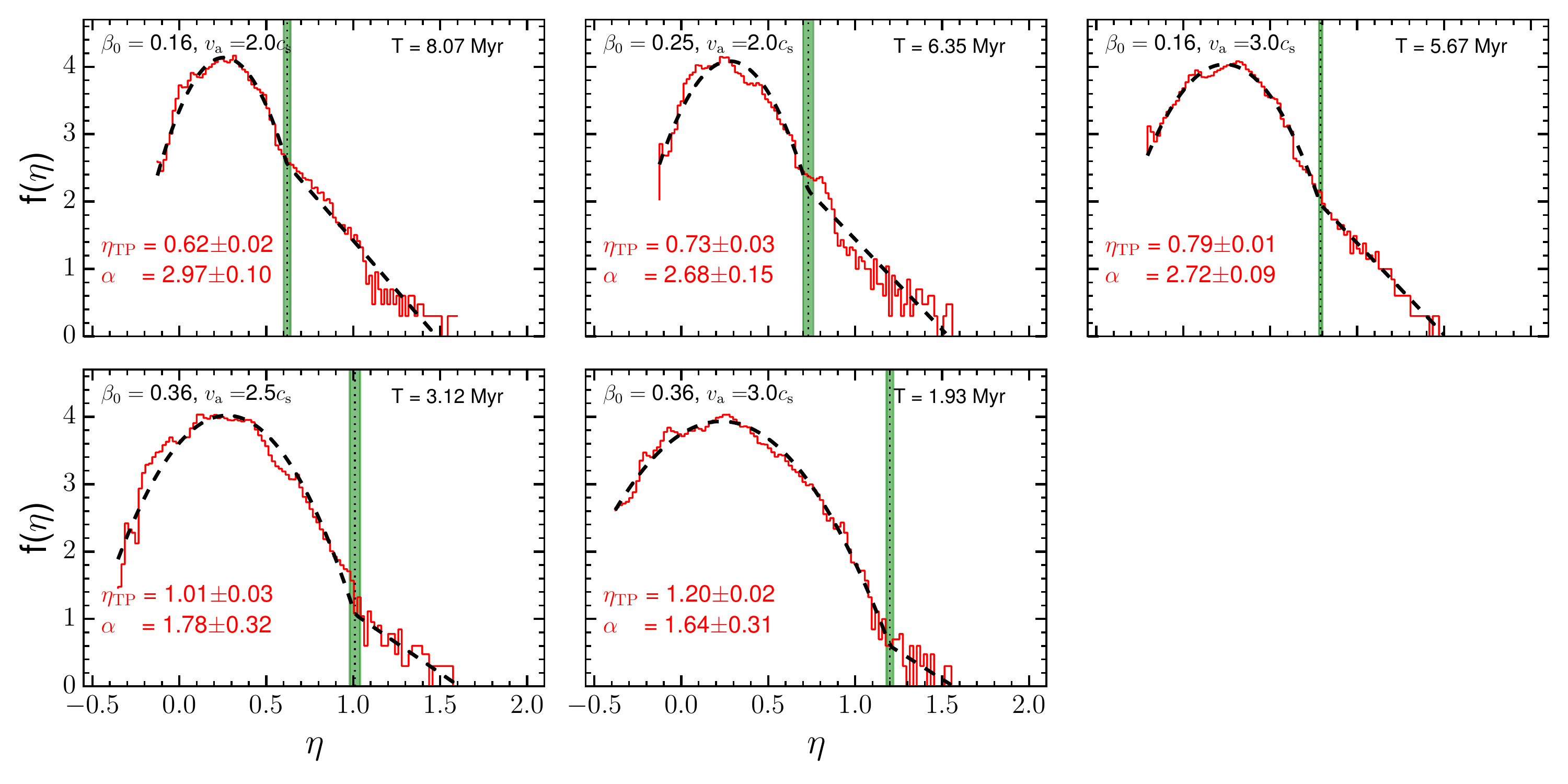}
    \caption{The column density PDFs of simulated models of molecular clouds with different initial conditions. The initial plasma $\beta_0$ and the turbulence amplitude $v_{\rm a}$ is specified on the top left of each panel. Each column density PDF is fitted with Equation (\ref{LGPLfit}). The vertical black dotted line marks the logarithmic column density value (i.e., transition point $\eta_{\rm TP}$) at which the power-law begins. The green shaded region represents the standard deviation of the transitional column density value. The best-fit parameters ($\alpha,\, \eta_{\rm{TP}}$) are obtained using the MCMC method in python. \label{fig:NPDFS}}

\end{center}
\end{figure*}
 
% \begin{figure*}
% \gridline{\fig{./figures/b16_v20.eps}{0.34\textwidth}{}
%           \fig{./figures/b25_v20.eps}{0.34\textwidth}{}
%           \fig{./figures/b16_v30.eps}{0.34\textwidth}{}\vspace{-.9cm}} 
% \gridline{\fig{./figures/b25_v30.eps}{0.34\textwidth}{}
%           \fig{./figures/b36_v25.eps}{0.34\textwidth}{}
%           \fig{./figures/b36_v30.eps}{0.34\textwidth}{}
%           }
% %\gridline{\fig{KT_Eri-eps-converted-to.pdf}{0.3\textwidth}{(f)}}
% \caption{The column density PDFs of simulated models of molecular clouds with different initial conditions. The initial magnetic field strength $\beta$ and the turbulence amplitude $v_{\rm a}$ is specified on the top left of each panel. Each column density PDF is fitted with Equation (\ref{LGPLfit}). The vertical black dotted line marks the logarithmic column density value (i.e., transition point $\eta_{_{\rm TP}}$) at which the power-law begins. The green shaded region represents the standard deviation of the transitional column density value. The best-fit parameters ($\alpha,\, \mu, \rm{TP}, \rm{and} \,\sigma)$are obtained using both the MCMC and curve-fit method in python. (Need to decide if we want to mention the chi-square values and even show the curve-fit values)\label{fig:NPDFS}}
% \end{figure*}

We study the time evolution of the column density PDFs for five different models using three-dimensional magnetohydrodynamic (MHD) simulations including self-gravity and ambipolar diffusion. The numerical setup is similar to the ones previously used in \cite{kud07,kud11} and \cite{aud18}, and we run with a  
%The only modification in our current simulations is the improved resolution in both the $x-$ and $-y$ direction. The 
number of grid points in each direction $(N_{x},N_{y}, N_{z}) = (512,512,20)$.

We consider models with a subcritical initial mass-to-flux ratio so that the magnetic field strength is dynamically important. The initial turbulent flow field causes the PDF to have a predominantly lognormal shape. However, as the cloud evolves, it forms compressed regions due to the large-scale flow and develops pockets of high column density. It then rebounds and shows oscillations. With each successive compression, more regions with high column density develop and cause a gradual widening of the width of the lognormal, even though the turbulence is decaying (see also \cite{war14,tas10,aud18}).

We follow the time evolution of the column density PDF for each model with different initial conditions. The PDF evolves over time from primarily having a lognormal shape at early times to developing a power-law tail (final time $\sim 1 \, \rm {Myr}$) when gravity dominates. The power-law develops after several oscillations, as the local pockets of higher column density become supercritical and go into a runaway collapse.

\subsection{Numerical Parameters}
The initial state has a uniform density in $x,y$ and is stratified in the $z-$direction
with a scale length
$H_{0} = c_{\rm {s0}}/\sqrt{2 \pi G \rho_{0}}$, where $c_{\rm s0}$ and $\rho_0$ are the isothermal sound speed and density at the midplane $z=0$. For more details of the initial setup, see \citet{kud07,kud11,aud18}. We choose $H_0$, $c_{\rm {s0}}$ and $\rho_{0}$ as units of length, velocity and density respectively. This gives the unit of time $t_{0} \equiv H_{0}/c_{\rm {s0}}$. The ratio of the initial gas to magnetic pressure at $z=0$ is 
\begin{equation}\label{beta}
\beta_{0} = \frac{8 \pi c_{\rm s0}^{2} \rho_{0}}{B_{0}^{2}},
\end{equation}
where $B_0$ is the initially uniform vertical magnetic field. We input Gaussian random velocity fluctuations of amplitude $v_{a}$ for each of the $x-$ and $y-$components of velocity, 
%at each grid points ($v_{x} = v_{a}R_{m}(x,y)$, $v_{y} = v_{a}R_{m}(x,y)$, $v_{z}$=0), where $R_{m}$ is random number with a spectrum 
in which the Fourier spectrum is $v_{k}^{2} \propto k^{-4}$. Appropriate choices of $\rho_{0}$ and $c_{s0}$ lead to dimensional values of standard quantities. 
For example, if $n_{0} \equiv \rho_{0}/m_n = 10^{4}$ cm$^{-3}$ where $m_n = 2.33 \times 1.67 \times 10^{-24}$ g, and $c_{\rm s0} = 0.2$ km s$^{-1}$, we get $H_{0} \simeq 0.05 $ pc and $t_{0} \simeq 2.5 \times 10^{5}\,\rm {yrs}$. If $\beta_{0} = 0.16$, Equation (\ref{beta}) yields $B_{0} \simeq$ 50 $\mu \rm G$. The initial column density is $\Sigma_{0} = \rho_{0} H_{0} \simeq 6 \times 10^{-3}$ g cm$^{-2}$, therefore the number column density is $N_{0} \equiv \Sigma_{0}/m_{n} \simeq 1.5 \times 10^{21}$ cm$^{-2}$.
%rizes the results for all the models and for different initial conditions.

\begin{table*}
	\centering
	\caption{Model and Fit Parameters}
	\label{Parameter}
	\begin{tabular}{llllcccc} % four columns, alignment for each
		\hline
		\hline
		Model & $v_{\rm {a}} / c_{\rm {s}}$  & $\beta_0$ & $\mathcal{M}^2_{\rm 0} \beta_0$ & $\eta_{\rm TP}$ & $|\alpha|$ & $\mu $ & $\sigma$ \\
		\hline
		%V0 &  0.25 & $k^{-4}$ & 0.01 & 149   \\
		T1 &  2.0  & 0.16  & 1.28 & 0.62 $\pm$ 0.02 & 3.0 $\pm$ 0.1 & 0.59 $\pm$ 0.01& 0.31 $\pm$ 0.01 \\
		T2 &  2.0  & 0.25  & 2.00 & 0.74 $\pm$ 0.03 & 2.7 $\pm$ 0.2 & 0.64 $\pm$ 0.02 & 0.35 $\pm$ 0.01 \\
		T3 &  3.0  & 0.16  & 2.88 & 0.79 $\pm$ 0.02 & 2.7 $\pm$ 0.1 & 0.55 $\pm$ 0.01 & 0.41 $\pm$ 0.01\\
% 		T4 &  3.5  & 0.16  & 3.92 &     \\
%		T4 &  3.0  & 0.25  & 4.50 & 0.80 $\pm$ 0.03 & 2.6 $\pm$ 0.2 & 0.65 $\pm$ 0.02 & 0.38 $\pm$ 0.01\\
		T4 &  2.5  & 0.36  & 4.50 & 1.01 $\pm$ 0.03 & 1.8 $\pm$ 0.3 & 0.63 $\pm$ 0.01& 0.46 $\pm$ 0.01\\
		T5 &  3.0  & 0.36  & 6.48 & 1.20 $\pm$ 0.02 & 1.6 $\pm$ 0.3 & 0.53 $\pm$ 0.01 & 0.57 $\pm$ 0.01\\

 		%K4 &  0.25 & $k^{0}$  & 10.0 & 69.7  \\
		\hline
	\end{tabular}
	\tablecomments{Fit parameters for the piecewise lognormal and power-law function are $\alpha,\, \sigma,\, \mu, \,\rm{and} \, \eta_{\rm{TP}}$. $\beta_{0} $ is the initial ratio of thermal to magnetic pressure at $z = 0$, $v_a$ is the amplitude of the initial velocity fluctuation and $\mathcal{M}_{\rm 0}$ is the sonic Mach number.}\label{tab:summary}
\end{table*}

\subsection{Fitting Functions}
In order to characterise the shape of the PDFs including the transition from the lognormal to power-law tail we consider two fitting functions: a purely lognormal function and a piecewise function that is a combination of a lognormal and a power law. If $f(\eta)$ is the PDF, 
the lognormal model is 
\begin{equation}\label{LGfit}
 f(\eta)_{\rm {LN}} = \log\left[ A \frac{1}{ \sqrt{2 \pi \sigma^{2}}} \exp \left(- \frac{(2.3 \eta - \mu)^2}{2\sigma^{2}} \right) \right] ,
 \end{equation}
where $\eta = \log (\Sigma / \Sigma_{0})$, $A= \ln (10) \times N_{\rm {total}}\footnote{Total length of the sample space of $\eta$ from the simulations, i.e., $512 \times 512$.}\times \Delta  \log (\Sigma / \Sigma_{0}) $ is the normalization constant, $\mu$ is the mean and $\sigma$ is the standard deviation. The data is binned with a uniform spacing of $\Delta \log (\Sigma / \Sigma_{0}) \simeq 0.02$. 
%We take the log of the lognormal function as we are fitting logarithmic data.
For the piecewise function \citep[see also][]{mye15,pok16} we consider a combination of a lognormal and a power law:
\begin{eqnarray}\label{LGPLfit}
 f(\eta)_{\rm LNPL} =& f(\eta)_{\rm {LN}}\, ,&\rm {if}\, \, \eta \le \eta_{\rm TP},  \nonumber \\
  =& f(\eta_{\rm TP})_{\rm {LN}}+ \alpha \eta   \, ,& \rm {if} 
  \, \, \eta > \eta_{\rm TP} ,
\end{eqnarray}
where $\alpha$ is the index of the power law and $\eta_{\rm TP} \equiv\log (\Sigma_{\rm TP} / \Sigma_{0)})$ is the logarithmic value of the transition column density. 
Thus, we have two fitting functions and four free parameters: $\mu ,\, \sigma ,\, \eta_{\rm TP},$ and $\alpha$. 

We use this piecewise four-parameter function rather than a mathematically simpler three-parameter continuous function, for example the modified lognormal power law distribution \citep{bas15}, since it clearly identifies a transition point $\eta_{\rm TP}$. We are then also using the same means to identify the transition point as used in observational analyses like \citet{pok16}.

% \begin{table*}
% 	\centering
% 	\caption{Model and Fit Parameters}
% 	\label{Parameter}
% 	\begin{tabular}{llllcccc} % four columns, alignment for each
% 		\hline
% 		\hline
% 		Model & $v_{\rm {a}} / c_{\rm {s}}$  & $\beta_0$ & $\mathcal{M}^2_{\rm 0} \beta_0$ & $\eta_{\rm TP}$ & $|\alpha|$ & $\mu $ & $\sigma$ \\
% 		\hline
% 		%V0 &  0.25 & $k^{-4}$ & 0.01 & 149   \\
% 		T1 &  2.0  & 0.16  & 1.28 & 0.62 $\pm$ 0.02 & 3.0 $\pm$ 0.1 & 0.59 $\pm$ 0.01& 0.31 $\pm$ 0.01 \\
% 		T2 &  2.0  & 0.25  & 2.00 & 0.74 $\pm$ 0.03 & 2.7 $\pm$ 0.2 & 0.64 $\pm$ 0.02 & 0.35 $\pm$ 0.01 \\
% 		T3 &  3.0  & 0.16  & 2.88 & 0.79 $\pm$ 0.02 & 2.7 $\pm$ 0.1 & 0.55 $\pm$ 0.01 & 0.41 $\pm$ 0.01\\
% % 		T4 &  3.5  & 0.16  & 3.92 &     \\
% %		T4 &  3.0  & 0.25  & 4.50 & 0.80 $\pm$ 0.03 & 2.6 $\pm$ 0.2 & 0.65 $\pm$ 0.02 & 0.38 $\pm$ 0.01\\
% 		T4 &  2.5  & 0.36  & 4.50 & 1.01 $\pm$ 0.03 & 1.8 $\pm$ 0.3 & 0.63 $\pm$ 0.01& 0.46 $\pm$ 0.01\\
% 		T5 &  3.0  & 0.36  & 6.48 & 1.20 $\pm$ 0.02 & 1.6 $\pm$ 0.3 & 0.53 $\pm$ 0.01 & 0.57 $\pm$ 0.01\\

%  		%K4 &  0.25 & $k^{0}$  & 10.0 & 69.7  \\
% 		\hline
% 	\end{tabular}
% 	\tablecomments{Fit parameters for the piecewise lognormal and power-law function are $\alpha,\, \sigma,\, \mu, \,\rm{and} \, \eta_{\rm{TP}}$. $\beta_{0} $ is the initial ratio of thermal to magnetic pressure at $z = 0$, $v_a$ is the amplitude of the initial velocity fluctuation and $\mathcal{M}_{\rm 0}$ is the sonic Mach number.}\label{tab:summary}
% \end{table*}
%in each of the $x-$ and $y-$ directions,
\subsection{Fitting the Simulation Data}
The value of the fit parameters is essential to characterise the shape of the column density PDFs. However, the column density PDFs are evolving in time with their shape changing from purely lognormal to a hybrid function. Thus it is essential to have a robust fitter that can capture the transition and identify $\eta_{\rm TP}$. We fit the column density PDFs at different times using $f(\eta)_{\rm {LN}}$ (Equation \ref{LGfit}) and $f(\eta)_{\rm LNPL}$ (Equation \ref{LGPLfit}) and compute the resulting $\chi^{2}$ values. We accept $f(\eta)_{\rm LNPL}$ only when its $\chi^{2}$ value is less than 20\% of that of the simpler lognormal function $f(\eta)_{\rm {LN}}$ and the power-law index $\alpha <5$.  
For models that are fit with $f(\eta)_{\rm LNPL}$, we further use a Markov Chain Monte Carlo (MCMC) method \citep{van03} as a second fitter. This gives us more robust best-fitting values from the parameter space along with reliable uncertainties. We have used the PYTHON package Pymc for this purpose \citep{pat10}.

The power-law tail appears during the final stages of the simulations, primarily due to ambipolar-diffusion-driven gravitational contraction. All the free parameters evolve moderately, including the transitional column density $\eta_{\rm{TP}}$, which grows by $\approx 10 \%$ from its initial appearance until the final time step. For simplicity we only consider the column density PDFs at the end of the simulation when the maximum density has reached 100$\rho_{0}$. Runaway collapse has ensued in the high density regions at this time. 
While we cannot follow the PDFs into the protostellar phase in these simulations, we anticipate that the large-scale maps of the PDFs will remain largely the same.

% Hence, we need a robust fitter that can give the best-fitting values from the parameter space along with reliable uncertainties. For this, we used the Markov Chain Monte Carlo (MCMC) method (van Dyk 2003). We have followed the Metropolis-Hastings algorithm in which the samples are selected from an arbitrary ‘proposal’ distribution. These samples are kept or discarded according to the acceptance rule. The whole process is repeated until we get a ‘transition’ probability function so that the algorithm can transit from one set of parameter values to a more probable set. Based on the transition probability where the current point depends only on the previous point but yet can still span over the whole parameter space, an ergodic chain of positions in parameter space is formed, known as the Markov Chain. The Markov Chain samples from the posterior distribution ergodically assuming the detailed balance condition. MCMC estimates the expectation of a statistic in a complex model by doing simulations that randomly select from a Markov Chain.

Figure \ref{fig:NPDFS} shows the column density PDFs of five models with different initial conditions (i.e., $\beta_{0}$ and $v_{\rm{t0}}$) along with the best fit lognormal and power-law functions. The results are summarized in Table \ref{tab:summary}. The run time for each simulation is indicated on the top right of each plot. The best-fitting parameters $\alpha, \,\rm{and} \, \eta_{\rm{TP}}$ are also shown. The plots are arranged from top left to bottom right according to increasing values of $\mathcal {M}^2_{\rm 0} \beta_0$. The black dotted line marks the column density value at which the lognormal PDF ends and the power-law tail begins. The green shaded region shows the standard deviation of the $\eta_{\rm{TP}}$ value obtained from the MCMC fit.     

The transition point $\eta_{\rm {TP}}$ shifts toward higher column density with increasing value of $\mathcal {M}^2_{\rm 0} \beta_0$. For example, it is minimum for model T1 ($\mathcal {M}^2_{\rm 0} \beta_0 =1.28$) and maximum for model T5 ($\mathcal {M}^2_{\rm 0} \beta_0 =6.48$) with $\eta_{\rm {TP}} = 0.62 \pm 0.02$ and $\eta_{\rm {TP}} = 1.20 \pm 0.02$, respectively. The value depends on the strength of the initial magnetic field and the amplitude of the velocity perturbation.

\section{Analytic Model} \label{sec:analytic}
We can understand the physical origin of the transition point with an analytic model in which the magnetic field is dynamically important. The cloud flattens along the mean magnetic field direction ($z$) and the subsequent evolution is primarily perpendicular to the magnetic field. The mass-to-flux ratio is subcritical until ambipolar diffusion creates supercritical pockets that are prone to collapse. Turbulence causes the creation of locally compressed regions that have a pressure balance between magnetic and ram pressure. This results in the formation of magnetic ribbons \cite[][see also \cite{kud14}]{aud16}. The cloud is stratified along the $z$-direction with compression along the $x-y$ plane. We simplify the analysis and assume that the thermal pressure is negligible compared to the magnetic pressure and the ram pressure of the flow. 
%Using similar argument as in \cite{aud16}, we consider a force-balance state. 
The pressure due to the magnetic field $B$ upon compression balances the initial pressure due to the background magnetic field $B_{0}$ and the external ram pressure in the $x-y$ direction:

\begin{equation}\label{forcebalance}
H\frac{B^2}{8 \pi} = H_{0}\left(\rho_{0}v_{\rm {t0}}^{2} + \frac{B_{0}^2}{8 \pi}\right),
\end{equation}
where $v_{\rm {t0}} = \sqrt{2}v_{a}$ is the nonlinear flow speed.
This results in a quasi-equilibrium state as compression ceases and oscillations begin. The gas has already settled into a hydrostatic equilibrium along the $z-$ direction and the cloud has a half-thickness

\begin{equation}\label{scaleheight}
H= \frac{c_{\rm s}}{\sqrt{2 \pi G \rho}}
\end{equation}
\citep{spi42}. Integrating the density along the scale height $H$ in the $z-$direction gives the column density 
\begin{equation}\label{columndensity}
\Sigma = 2  \rho H.
\end{equation}

% The initial density compression is very high \citep[see figure 1 in][]{kud08} but the subcritical mass-to-flux ratio results in a strong rebound. Subsequent oscillations are not as strong (due to the decaying turbulence) but with each oscillation there is some loss of magnetic flux due to ambipolar diffusion. Thus, after a few oscillations the mass-to-flux ratio of the compressed state is greater than the initial value, i.e.,
% Considering that the ambipolar diffusion time is much longer than the compression time the cloud is nearly flux frozen during the compression, i.e., 
% \begin{equation}\label{flux-frozen}
% \frac{\Sigma}{B} = a_1  \frac{\Sigma_{0}}{B_0},
% \end{equation}
% {\bf where $a \gtrsim 1$. Using Equations (\ref{scaleheight}),(\ref{columndensity}) and (\ref{flux-frozen}) in Equation (\ref{forcebalance}), and allowing for some decay of the initial turbulence so that the turbulent amplitude is $v^2_{\rm t} = a_2\,v^2_{\rm t0}$ where $a_2 \lesssim 1$, we find
The initial density compression is very high \citep[see figure 1 in][]{kud08} but the subcritical mass-to-flux ratio results in a strong rebound.
We consider that the cloud is nearly flux frozen during its initial compression, i.e., $B/\Sigma = \rm constant$, as the ambipolar diffusion time is much longer than the compression time. Using (\ref{scaleheight}) and (\ref{columndensity}) in Equation (\ref{forcebalance}) along with the flux frozen condition we find 
\begin{equation}\label{FB2} 
\frac{\Sigma}{\Sigma_{0}} = \left[ v_{\rm {t0}}^2 \left(\frac{8 \pi \rho_{0}}{B_{0}^2} \right) + 1\right].
\end{equation}
% where $a = a_1a_2$ is a correction factor of order unity that contains uncertainties about
% the flux loss and turbulent decay. 
The force balance of Equation (\ref{FB2}) gives a critical column density that we denote as $\Sigma_{\rm {TP}}$.
However, subsequent oscillations are not as strong due to decay of the initial turbulence amplitude $v_{\rm t0}$ and loss of magnetic flux due to ambipolar diffusion.
Allowing for such variations we rewrite Equation (\ref{FB2}) in terms of the sonic Mach number $\mathcal{M}_{\rm 0} = v_{\rm t0}/c_{\rm s}$, and plasma $\beta_{0}$ as
\begin{equation}\label{eq:FB3}
\frac{\Sigma_{\rm TP}}{\Sigma_{0}} = a (\mathcal{M}^2_{\rm 0} \beta_{0} + 1),
\end{equation}
% Equation (\ref{eq:FB3}) gives the variation of $\Sigma_{\rm TP}$ as a function of the initial condition $\mathcal{M}^{2} \beta_{0}$.
where $a$ is a correction factor of order unity that contains uncertainties about the flux loss and turbulent decay.

Here $\mathcal{M}^2_{\rm 0} \beta_{0} \equiv 2 \mathcal{M}^2_{\rm A0} $, where $\mathcal{M_{\rm A0}} = v_{\rm t0}/v_{\rm A0}$ is the initial Alfv{\'e}nic Mach number and $v_{\rm A0} = B_{0}/ \sqrt{(4 \pi \rho_{0})} $ is the initial Alfv{\'e}n speed in the midplane. 
% For the square of the initial Alfv{\'e}n velocity $v_{\rm A0}^{2} = B_{0}^2/ (4 \pi \rho_{0}) $, Equation (\ref{eq:FB3})  can be expressed in terms of the Alfv{\'e}nic Mach number $\mathcal{M_{\rm A0}} = v_{\rm t0}/v_{\rm A0}$ as 
% \begin{equation}\label{eq:FB4}
% \frac{\Sigma_{\rm TP}}{\Sigma_{0}} = a (2\mathcal{M_{\rm {A0}}}^2 + 1).
% \end{equation}

% Thus the transition is related to the initial parameters ($\mathcal{M}$ and $\beta_{0}$) of the turbulence-enhanced-ambipolar diffusion driven star formation scenario. 
% The parameters $\mathcal{M}$ and $\beta_{0}$, depicts the initial condition of the turbulence-enhanced-ambipolar diffusion driven star formation scenario. 

% The above equation gives the normalized column density of the compressed regions set by the standoff between the ram pressure and the magnetic pressure. 

% The compressed regions with enhanced column density $\Sigma$ eventually collapse due to gravitationally driven ambipolar diffusion. Equation (\ref{FB2}) can also we written as 
% \begin{equation}
%     \frac{\Sigma}{\Sigma_{0}} = a\left[2 \left(\frac{v_{\rm t0}}{v_{\rm A}}\right)^2 + 1\right],
% \end{equation}
% where $v_{\rm A0}^{2} = B_{0}^2/ (4 \pi \rho_{0}) $ is the square of the initial Alfv$`$en velocity.

\subsection{Physical Interpretation of $\Sigma_{\rm TP}$}\label{fitting}
\begin{figure}
\begin{center}
	\includegraphics[height=6cm,width=8.5cm,trim=3mm 5mm 3mm 0mm, clip=true]{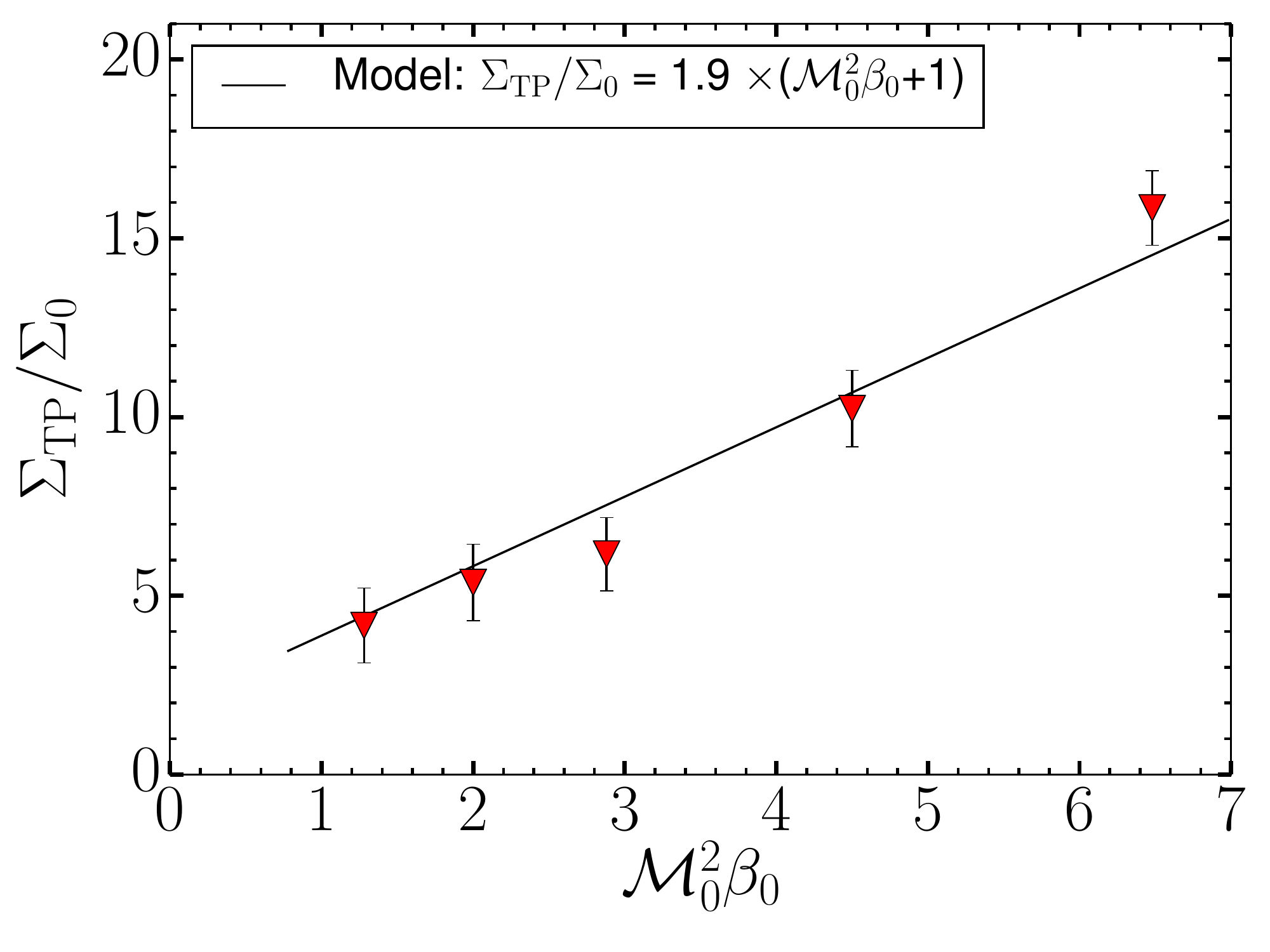}
    \caption{The normalized transition column density $(\Sigma_{\rm {TP}}/\Sigma_{0}) =10^{\eta_{\rm {TP}}}$ for different values of initial $\mathcal{M}^2_{\rm 0} \beta_{\rm 0}$ obtained from the simulations. The black line is the theoretical model (Equation \ref{eq:FB3}) with the best fit value of $a = 1.9$.  }
    \label{fig:theoryfit}
\end{center}
	\end{figure}

Figure \ref{fig:theoryfit} shows the variation of the normalized transitional column density $(\Sigma_{\rm {TP}}/\Sigma_{0})$ for simulated models with different initial values of $\mathcal{M}^2_{\rm 0} \beta_{\rm 0}$. We fit the analytic expression (Equation \ref{eq:FB3}) to the simulation data and get a best fit value of $a = 1.9$. We find a good agreement between the simulation data and our analytic model. 

The transition from the lognormal to the power-law tail signifies both structural and morphological changes. It marks a transition from the ambient subcritical turbulent background ($\Sigma < \Sigma_{\rm {TP}}$) to a compressed denser region where $\Sigma > \Sigma_{\rm {TP}}$. Due to ambipolar diffusion (ion-neutral drift) the force balance between the ram pressure and the magnetic field gradually relaxes. There is a gradual loss of magnetic flux as the neutrals diffuse past the ions with each successive oscillation. The density is enhanced after each compression resulting in an increase of mass-to-flux ratio. The transitional column density $\Sigma_{\rm TP}$ defines this cutoff beyond which the mass-to-flux ratio becomes critical. For $\Sigma > \Sigma_{\rm {TP}}$ gravity becomes increasingly important and the power-law tail emerges.  

 Furthermore, with increased strength of the Alfv{\'e}nic Mach number $\mathcal{M}_{\rm A0}$ the initial compression is much stronger and it results in a higher density. This causes $\Sigma_{\rm TP}$ to shift towards higher values with increasing strength of $2\mathcal{M}_{\rm A0}^2$ .

	\section{Discussion} \label{sec:discussion}
The shape of the column density PDF is an imprint of the initial conditions in a star-forming molecular cloud. It holds the key to finding the link between the structural properties and the ambient conditions that trigger star formation in molecular clouds.
% The width of the lognormal, the power-law slope,  and the transitional column density are characterised by the combined action of turbulence, magnetic field and gravity. 
However, it is difficult to detect low column density material (the lognormal part) using dust emission and extinction measurement because of observational biases \citep{lom15}. The lognormal peak is sometimes considered an artefact arising due to data incompleteness \citep{alv17} or undetectable due to insufficient sampling or limited field of view \citep{bas19}. The more robust observational quantities are the characteristic break $\Sigma_{\rm TP}$ in the PDF and the power-law slope ($\alpha$), as these are less affected by such constraints \citep{lom15}.
% \cite{aud18, col12} showed that magnetic fields affects the power-law slope ($\alpha$), but analytic expression to quantify it.  

Our model shows that $\Sigma_{\rm TP}$ is a measure of the initial Alfv{\'e}nic Mach number $\mathcal{M}_{\rm A0}$. Observations of $\Sigma_{\rm TP}$ can be used to infer the
value of $\mathcal{M}_{\rm A0} \propto \mathcal{M}_0 \beta_0^{1/2}$. This could in 
principle also lead to an estimate of the initial normalized mass-to-flux ratio $\mu_0$, where $\mu_0^2 \simeq \beta_{0}$ \citep[see][]{kud07}, if the initial Mach number $\mathcal{M}_0$ can be estimated. This is potentially important since a direct measurement of the magnetic field strength using the Zeeman effect is difficult \citep{cru12}. Indirect probes of the magnetic field such as dust polarization \citep{hoa08}, spectroscopic methods \citep{aud19}, and Faraday rotation \citep{wol04} also have their limitations. An estimate of $\mathcal{M}_0$ can in principle be made from the observed width $\sigma$ of the lognormal PDF by developing an analytic/empirical relation that captures the increase of $\sigma$ with the increasing strength of $\mathcal{M}^2_{\rm 0} \beta_{0}$ (see Table \ref{Parameter}). Such an analysis can be pursued in future work.

\section{Conclusion} \label{sec:conclusion}
Our key findings are:
\begin{itemize}
    
    \item The transitional column density $\Sigma_{\rm {TP}}$ represents a transition from a turbulent magnetically-dominated background ($\Sigma< \Sigma_{\rm {TP}}$) having lognormal shape to a dense region ($\Sigma> \Sigma_{\rm {TP}}$) with a power-law tail where gravity is dominant.
    
    \item $\Sigma_{\rm {TP}}$ marks the boundary between regions with subcritical (magnetically dominated) and supercritical (weak magnetic field) mass-to-flux ratio in a star-forming molecular cloud.
    
    \item $\Sigma_{\rm {TP}}$ depends on the initial velocity perturbation (sonic Mach number $\mathcal{M}_{0}$) and the magnetic field strength (plasma $\beta_{0}$). Alternatively, it is a measure of the initial Alfv{\'e}nic Mach number and increases with the increasing strength of $2\mathcal{M}^2_{\rm A0}$.

\end{itemize}

\section*{Acknowledgements}
SA acknowledges support from an ASIAA Postdoctoral Fellowship. Computations were carried out using facilities of SHARCNET. SB is supported by a Discovery Grant from NSERC.


\begin{thebibliography}{}
\expandafter\ifx\csname natexlab\endcsname\relax\def\natexlab#1{#1}\fi
\providecommand{\url}[1]{\href{#1}{#1}}

\bibitem[{{Alves} {et~al.}(2014){Alves}, {Lombardi}, \& {Lada}}]{alv14}
{Alves}, J., {Lombardi}, M., \& {Lada}, C.~J. 2014, \aap, 565, A18

\bibitem[{{Alves} {et~al.}(2017){Alves}, {Lombardi}, \& {Lada}}]{alv17}
---. 2017, \aap, 606, L2

\bibitem[{{Auddy} {et~al.}(2016){Auddy}, {Basu}, \& {Kudoh}}]{aud16}
{Auddy}, S., {Basu}, S., \& {Kudoh}, T. 2016, \apj, 831, 46

\bibitem[{{Auddy} {et~al.}(2018){Auddy}, {Basu}, \& {Kudoh}}]{aud18}
---. 2018, \mnras, 474, 400

\bibitem[{{Auddy} {et~al.}(2019){Auddy}, {Myers}, {Basu}, {Harju}, {Pineda}, \&
  {Friesen}}]{aud19}
{Auddy}, S., {Myers}, P.~C., {Basu}, S., {et~al.} 2019, \apj, 872, 207

\bibitem[{{Ballesteros-Paredes} {et~al.}(2011){Ballesteros-Paredes},
  {V{\'a}zquez-Semadeni}, {Gazol}, {Hartmann}, {Heitsch}, \&
  {Col{\'{\i}}n}}]{par11}
{Ballesteros-Paredes}, J., {V{\'a}zquez-Semadeni}, E., {Gazol}, A., {et~al.}
  2011, \mnras, 416, 1436
  
\bibitem[Basu et al.(2015)]{bas15} Basu, S., Gil, M., \& Auddy, S.\ 2015, \mnras, 449, 2413

\bibitem[{{Burkhart}(2018)}]{bur18}
{Burkhart}, B. 2018, \apj, 863, 118

\bibitem[{{Burkhart} {et~al.}(2015{\natexlab{a}}){Burkhart}, {Collins}, \&
  {Lazarian}}]{bur15}
{Burkhart}, B., {Collins}, D.~C., \& {Lazarian}, A. 2015{\natexlab{a}}, \apj,
  808, 48

\bibitem[{{Burkhart} {et~al.}(2015{\natexlab{b}}){Burkhart}, {Lee}, {Murray},
  \& {Stanimirovi{\'c}}}]{bur15b}
{Burkhart}, B., {Lee}, M.-Y., {Murray}, C.~E., \& {Stanimirovi{\'c}}, S.
  2015{\natexlab{b}}, \apjl, 811, L28

\bibitem[{{Burkhart} {et~al.}(2017){Burkhart}, {Stalpes}, \& {Collins}}]{bur17}
{Burkhart}, B., {Stalpes}, K., \& {Collins}, D.~C. 2017, \apjl, 834, L1

\bibitem[Blitz(1991)]{bli91} Blitz, L.\ 1991, NATO Advanced Science Institutes (ASI) Series C, 3

\bibitem[{{Collins} {et~al.}(2012){Collins}, {Kritsuk}, {Padoan}, {Li}, {Xu},
  {Ustyugov}, \& {Norman}}]{col12}
{Collins}, D.~C., {Kritsuk}, A.~G., {Padoan}, P., {et~al.} 2012, \apj, 750, 13

\bibitem[{{Crutcher}(2012)}]{cru12}
{Crutcher}, R.~M. 2012, \araa, 50, 29

\bibitem[{{Federrath} \& {Klessen}(2013)}]{fed13}
{Federrath}, C., \& {Klessen}, R.~S. 2013, \apj, 763, 51

\bibitem[{{Federrath} {et~al.}(2008){Federrath}, {Klessen}, \&
  {Schmidt}}]{fed08}
{Federrath}, C., {Klessen}, R.~S., \& {Schmidt}, W. 2008, \apjl, 688, L79

\bibitem[{{Goldsmith} {et~al.}(2008){Goldsmith}, {Heyer}, {Narayanan}, {Snell},
  {Li}, \& {Brunt}}]{gol08}
{Goldsmith}, P.~F., {Heyer}, M., {Narayanan}, G., {et~al.} 2008, \apj, 680, 428

\bibitem[{{Hoang} \& {Lazarian}(2008)}]{hoa08}
{Hoang}, T., \& {Lazarian}, A. 2008, \mnras, 388, 117

\bibitem[{{Imara} \& {Burkhart}(2016)}]{ima16}
{Imara}, N., \& {Burkhart}, B. 2016, \apj, 829, 102

\bibitem[{{Kainulainen} {et~al.}(2009){Kainulainen}, {Beuther}, {Henning}, \&
  {Plume}}]{kai09}
{Kainulainen}, J., {Beuther}, H., {Henning}, T., \& {Plume}, R. 2009, \aap,
  508, L35

\bibitem[{{K{\"o}nyves} {et~al.}(2015){K{\"o}nyves}, {Andr{\'e}},
  {Men'shchikov}, {Palmeirim}, {Arzoumanian}, {Schneider}, {Roy}, {Didelon},
  {Maury}, {Shimajiri}, {Di Francesco}, {Bontemps}, {Peretto}, {Benedettini},
  {Bernard}, {Elia}, {Griffin}, {Hill}, {Kirk}, {Ladjelate}, {Marsh}, {Martin},
  {Motte}, {Nguy{\^e}n Luong}, {Pezzuto}, {Roussel}, {Rygl}, {Sadavoy},
  {Schisano}, {Spinoglio}, {Ward-Thompson}, \& {White}}]{kon15}
{K{\"o}nyves}, V., {Andr{\'e}}, P., {Men'shchikov}, A., {et~al.} 2015, \aap,
  584, A91

\bibitem[{{K{\"o}rtgen} {et~al.}(2019){K{\"o}rtgen}, {Federrath}, \&
  {Banerjee}}]{bas19}
{K{\"o}rtgen}, B., {Federrath}, C., \& {Banerjee}, R. 2019, \mnras, 482, 5233

\bibitem[{{Kritsuk} {et~al.}(2011){Kritsuk}, {Norman}, \& {Wagner}}]{kri11}
{Kritsuk}, A.~G., {Norman}, M.~L., \& {Wagner}, R. 2011, \apjl, 727, L20

\bibitem[{{Kudoh} \& {Basu}(2008)}]{kud08}
{Kudoh}, T., \& {Basu}, S. 2008, \apjl, 679, L97

\bibitem[{{Kudoh} \& {Basu}(2011)}]{kud11}
---. 2011, \apj, 728, 123

\bibitem[{{Kudoh} \& {Basu}(2014)}]{kud14}
---. 2014, \apj, 794, 127

\bibitem[{{Kudoh} {et~al.}(2007){Kudoh}, {Basu}, {Ogata}, \& {Yabe}}]{kud07}
{Kudoh}, T., {Basu}, S., {Ogata}, Y., \& {Yabe}, T. 2007, \mnras, 380, 499

\bibitem[{{Lombardi} {et~al.}(2015){Lombardi}, {Alves}, \& {Lada}}]{lom15}
{Lombardi}, M., {Alves}, J., \& {Lada}, C.~J. 2015, \aap, 576, L1

\bibitem[{{Molina} {et~al.}(2012){Molina}, {Glover}, {Federrath}, \&
  {Klessen}}]{mol12}
{Molina}, F.~Z., {Glover}, S.~C.~O., {Federrath}, C., \& {Klessen}, R.~S. 2012,
  \mnras, 423, 2680

\bibitem[{{Myers}(2015)}]{mye15}
{Myers}, P.~C. 2015, \apj, 806, 226

\bibitem[{{Padoan} {et~al.}(1997){Padoan}, {Jones}, \& {Nordlund}}]{pad97}
{Padoan}, P., {Jones}, B.~J.~T., \& {Nordlund}, {\AA}.~P. 1997, \apj, 474, 730

\bibitem[{{Passot} \& {V{\'a}zquez-Semadeni}(1998)}]{pas98}
{Passot}, T., \& {V{\'a}zquez-Semadeni}, E. 1998, \pre, 58, 4501

\bibitem[{Patil {et~al.}(2010)Patil, Huard, \& Fonnesbeck}]{pat10}
Patil, A., Huard, D., \& Fonnesbeck, C. 2010, Journal of Statistical Software,
  Articles, 35, 1.
\newblock \url{https://www.jstatsoft.org/v035/i04}

\bibitem[{{Pokhrel} {et~al.}(2016){Pokhrel}, {Gutermuth}, {Ali}, {Megeath},
  {Pipher}, {Myers}, {Fischer}, {Henning}, {Wolk}, {Allen}, \& {Tobin}}]{pok16}
{Pokhrel}, R., {Gutermuth}, R., {Ali}, B., {et~al.} 2016, \mnras, 461, 22

\bibitem[{{Scalo} {et~al.}(1998){Scalo}, {V{\'a}zquez-Semadeni}, {Chappell}, \&
  {Passot}}]{sca98}
{Scalo}, J., {V{\'a}zquez-Semadeni}, E., {Chappell}, D., \& {Passot}, T. 1998,
  \apj, 504, 835

\bibitem[{{Schneider} {et~al.}(2013){Schneider}, {Andr{\'e}}, {K{\"o}nyves},
  {Bontemps}, {Motte}, {Federrath}, {Ward-Thompson}, {Arzoumanian},
  {Benedettini}, {Bressert}, {Didelon}, {Di Francesco}, {Griffin}, {Hennemann},
  {Hill}, {Palmeirim}, {Pezzuto}, {Peretto}, {Roy}, {Rygl}, {Spinoglio}, \&
  {White}}]{sch13}
{Schneider}, N., {Andr{\'e}}, P., {K{\"o}nyves}, V., {et~al.} 2013, \apjl, 766,
  L17

\bibitem[{{Schneider} {et~al.}(2015{\natexlab{a}}){Schneider}, {Csengeri},
  {Klessen}, {Tremblin}, {Ossenkopf}, {Peretto}, {Simon}, {Bontemps}, \&
  {Federrath}}]{sch15b}
{Schneider}, N., {Csengeri}, T., {Klessen}, R.~S., {et~al.} 2015{\natexlab{a}},
  \aap, 578, A29

\bibitem[{{Schneider} {et~al.}(2015{\natexlab{b}}){Schneider}, {Ossenkopf},
  {Csengeri}, {Klessen}, {Federrath}, {Tremblin}, {Girichidis}, {Bontemps}, \&
  {Andr{\'e}}}]{sch15}
{Schneider}, N., {Ossenkopf}, V., {Csengeri}, T., {et~al.} 2015{\natexlab{b}},
  \aap, 575, A79

\bibitem[{{Schneider} {et~al.}(2016){Schneider}, {Bontemps}, {Motte},
  {Ossenkopf}, {Klessen}, {Simon}, {Fechtenbaum}, {Herpin}, {Tremblin},
  {Csengeri}, {Myers}, {Hill}, {Cunningham}, \& {Federrath}}]{sch16}
{Schneider}, N., {Bontemps}, S., {Motte}, F., {et~al.} 2016, \aap, 587, A74

\bibitem[{{Spitzer}(1942)}]{spi42}
{Spitzer}, Jr., L. 1942, \apj, 95, 329

\bibitem[{{Tassis} {et~al.}(2010){Tassis}, {Christie}, {Urban}, {Pineda},
  {Mouschovias}, {Yorke}, \& {Martel}}]{tas10}
{Tassis}, K., {Christie}, D.~A., {Urban}, A., {et~al.} 2010, \mnras, 408, 1089

\bibitem[{{van Dyk}(2003)}]{van03}
{van Dyk}, D.~A. 2003, {Hierarchical models, data augmentation, and Markov
  chain Monte Carlo}, ed. E.~D. {Feigelson} \& G.~J. {Babu}, 41--56

\bibitem[{{Ward} {et~al.}(2014){Ward}, {Wadsley}, \& {Sills}}]{war14}
{Ward}, R.~L., {Wadsley}, J., \& {Sills}, A. 2014, \mnras, 445, 1575

\bibitem[{{Wolleben} \& {Reich}(2004)}]{wol04}
{Wolleben}, M., \& {Reich}, W. 2004, \aap, 427, 537



\end{thebibliography}
\end{document}